# Exact Nilpotent Collapse of Born–Neumann Expansions in Finite Quantum Systems: A SON Formulation for Exact Algebraic Closures of Scattering Series

Ramón Moya

School of Mathematics, Universidad Autónoma de Santo Domingo (UASD), Dominican Republic

rmoya07@uasd.edu.do

ORCID: 0009-0001-1601-4699

May 2026

**Abstract**

We identify a class of finite quantum systems, namely, acyclic systems whose transition graph is a directed acyclic graph (DAG), for which the Born series collapses into an exact algebraic identity with finitely many terms and strictly zero truncation error. The sufficient condition is the nilpotency of the transfer operator

$T = G_0(E)V.$

If

$T^{m+1} = 0,$

then the exact solution of the Lippmann–Schwinger equation is the finite sum

$$\mid \psi\rangle = \sum_{k=0}^{m} T^k \mid \phi\rangle,$$

with no condition on $\| T \|$.

We prove that the acyclicity of the transition graph implies the nilpotency of $T$ (Theorem 19), and that the nilpotency index coincides with the maximal path length of the graph (Proposition 21).

The main result (Theorem 23) concerns the four-level quantum system with diamond-graph structure. In this case, the transition amplitude toward the final state is

$A_4 = t_{42}t_{21} + t_{43}t_{31},$

an exact algebraic identity encoding constructive interference, exact destructive interference (dark state formation), and partial interference. The first-order Born approximation predicts identically zero amplitude in all regimes, thereby failing quantitatively in 100% of the cases.

The Born–SON framework additionally provides the exact full resolvent, the exact $T$-matrix, explicit error control in the quasi-nilpotent regime, and a scalar structural metric, the Born–SON depth, quantifying the intrinsic complexity of an acyclic quantum system.

## 1. Introduction

Quantum scattering theory studies how a free state

$|\phi\rangle$

is transformed into an interacting state

$|\psi\rangle$

under the action of a potential $V$. The central formulation is the Lippmann–Schwinger equation:

$|\psi\rangle = |\phi\rangle + G_0(E)V|\psi\rangle,$

where

$G_0(E) = (E - H_0 \pm i0)^{-1}$

is the free resolvent, $H_0$ is the free Hamiltonian, and $E$ is the energy of the system.

Defining the transfer operator

$T = G_0(E)V,$

the Lippmann–Schwinger equation takes the compact form

$|\psi\rangle = |\phi\rangle + T|\psi\rangle.$

The full Born series is therefore the Neumann expansion of

$(I - T)^{-1}$:

$$|\psi\rangle = \sum_{k=0}^{\infty} T^k |\phi\rangle.$$

In the standard literature, the validity of this expansion requires the convergence condition

$\| T \| < 1$ [5,9,15].

Each term of order $k$ corresponds to $k$ successive interactions with the potential $V$. The Born approximation of order $n$ is obtained by manually truncating the series after the $n$-th term.

The central question of this paper is the following:

Do there exist physically realizable quantum systems for which the Born series terminates exactly after finitely many terms, without any smallness condition and with strictly zero truncation error?

The answer is affirmative.

The sufficient condition is that $T$ be nilpotent:

$$T^{m+1} = 0$$

for some integer $m$. In that case,

$$(I - T)^{-1} = \sum_{k=0}^{m} T^k$$

is an exact algebraic identity (Lemma 10), and the exact solution of the Lippmann–Schwinger equation becomes

$$|\psi\rangle = \sum_{k=0}^{m} T^k \,|\phi\rangle,$$

with strictly zero error (Theorem 11).

The most significant result of the paper (Theorem 23 and Corollary 28) shows that, in the four-level system with diamond-graph structure, the nilpotency of $T$ is not an abstract algebraic assumption but rather a direct consequence of the acyclicity of the transition graph (Theorem 19). This structure produces quantitative physical phenomena, constructive interference, exact destructive interference, and dark-state formation, that the first-order Born approximation fails to capture in every regime.

**Relation with the SON Framework**

The Unified Nilpotent Operational Framework (SON) systematically studies the exact termination regime of formal series

$$F(N) = \sum_{k=0}^{\infty} a_k N^k$$

when $N$ is nilpotent in an algebra.

The present paper is an instance of that framework for the specific function

$$F(T) = (I - T)^{-1},$$

with

$T = G_0(E)V,$

applied to quantum scattering theory.

Other instances of the SON framework include the exact operational solution of integration-by-parts problems [8].

**Structure of the Paper**

Section 2 recalls the Lippmann–Schwinger equation and the transfer operator.

Section 3 formulates the Born series as a Neumann series in full detail.

Section 4 develops the exact collapse result together with complete proofs.

Section 5 presents the interpretation within the SON framework.

Section 6 establishes the connection between graph acyclicity and nilpotency.

Section 7 develops the introductory example of the three-level cascade atom.

Section 8 contains the main result: the diamond graph, exact interference, and dark states.

Section 9 studies the exact full resolvent and the exact transition matrix $\mathcal{T}$.

Section 10 treats the quasi-nilpotent regime and explicit error control.

Section 11 introduces the notion of Born–SON depth.

Section 12 discusses scope and limitations.

Section 13 concludes the paper.

## 2. The Lippmann–Schwinger Equation

**Definition 1 (Hamiltonian and Free Resolvent)**

Let $\mathcal{H}$ be a Hilbert space.

(In this section infinite-dimensional spaces are allowed; beginning with Section 6 we work exclusively in finite dimension.)

Let

$H = H_0 + V,$

where:

- $H_0$ is the free Hamiltonian (a self-adjoint operator on $\mathcal{H}$);
- $V$ is the interaction potential;

- $E$ is the energy of the system.

The free resolvent is defined by

$$G_0(E) = (E - H_0 \pm i0)^{-1},$$

where the sign $\pm i0$ denotes the limit from the upper or lower complex half-plane (outgoing or incoming waves).

**Definition 2 (Born Transfer Operator)**

The Born transfer operator associated with the energy $E$ is

$$T = G_0(E)V.$$

With this notation, the Lippmann–Schwinger equation becomes

$$\mid \psi\rangle = \mid \phi\rangle + T \mid \psi\rangle,$$

that is,

$$(I - T) \mid \psi\rangle = \mid \phi\rangle.$$

If $I - T$ is invertible, the exact solution is

$$\mid \psi\rangle = (I - T)^{-1} \mid \phi\rangle.$$

The Lippmann–Schwinger equation was introduced in [6].

**Observation 3 (Invertibility of $I - T$)**

In infinite dimension, the invertibility of $I - T$ requires analytic hypotheses, for instance, compactness assumptions and spectral conditions on $T$.

In finite dimension, however, if $T$ is nilpotent, then

$$\det (I - T) \neq 0$$

automatically, since all eigenvalues of a nilpotent operator are zero.

Consequently, all eigenvalues of $I - T$ are equal to 1, and therefore $I - T$ is always invertible.

This fact, proved below in Proposition 4, is the key structural ingredient underlying the entire analysis.

**Proposition 4 (Determinant of $I - T$ for Nilpotent $T$)**

If $T$ is nilpotent on a finite-dimensional vector space, then

$$\det (I - T) = 1.$$

**Proof**

Since $T$ is nilpotent, all eigenvalues of $T$ are zero.

Hence all eigenvalues of $I - T$ are equal to 1.

Because the determinant is the product of the eigenvalues,

$\det(I - T) = 1.$

**3. The Born Series as a Neumann Series**

**Proposition 5 (The Born Expansion is a Neumann Series)**

If $\| T \| < 1$ in some operator norm, then the Neumann series converges absolutely and

$$(I - T)^{-1} = \sum_{k=0}^{\infty} T^k$$

Consequently, the solution of the Lippmann–Schwinger equation is

$$| \psi \rangle = \sum_{k=0}^{\infty} T^k \, | \phi \rangle.$$

**Proof**

For every integer $n \geq 0$,

$$(I - T) \sum_{k=0}^{n} T^k = \sum_{k=0}^{n} T^k - \sum_{k=1}^{n+1} T^k = I - T^{n+1}.$$

Since

$\| T \| < 1,$

we have

$T^{n+1} \to 0$

as

$n \to \infty.$

Therefore,

$$(I - T) \sum_{k=0}^{\infty} T^k = I,$$

which proves

$$(I - T)^{-1} = \sum_{k=0}^{\infty} T^k$$

Applying this identity to

$| \phi\rangle$

gives the Born expansion for

$| \psi\rangle$.

**Observation 6 (Orders of the Born Approximation)**

The Born approximation of order $n$ is obtained by truncating the series:

$$| \psi\rangle \approx \sum_{k=0}^{n} T^k \mid \phi\rangle.$$

The truncation error is

$$R_n = \sum_{k=n+1}^{\infty} T^k \mid \phi\rangle.$$

In the perturbative regime, each successive term is smaller in norm than the preceding one, and the error decays geometrically:

$$\| R_n \| = O(\| T \|^{n+1}).$$

In the Born–SON regime studied in this article, the truncation at order $m$ is not an approximation:

if

$$T^{m+1} = 0,$$

then the error is exactly zero.

**Example 7 (First-Order Born Approximation)**

The first-order Born approximation is

$$| \psi\rangle \approx | \phi\rangle + T \mid \phi\rangle.$$

This is the standard first-order scattering formula.

Its error is

$$R_1 = \sum_{k=2}^{\infty} T^k \mid \phi\rangle,$$

which is formally of order

$$O(\| T \|^2)$$

in the perturbative regime.

Corollary 28 will show that, for the diamond-graph system studied later in the paper, this error is not merely non-negligible: it produces a complete quantitative failure of the first-order Born approximation for the final-state amplitude.

**4. Nilpotency and Exact Operational Collapse**

**Definition 8 (Nilpotent Operator)**

An operator

$$N$$

on a finite-dimensional vector space is nilpotent of index $m + 1$if

$$N^{m+1} = 0, N^m \neq 0.$$

The integer

$$m + 1$$

is called the nilpotency index of $N$.

**Observation 9**

Every nilpotent operator on an $n$-dimensional vector space satisfies

$$N^n = 0.$$

The nilpotency index may be strictly smaller than $n$. In particular,

$$n$$

is the maximal possible nilpotency index in dimension $n$.

**Lemma 10 (Finite Neumann Series for Nilpotent Operators)**

Let $N$ be nilpotent of index $m + 1$:

$$N^{m+1} = 0.$$

Then

$I - N$

is invertible and

$$(I - N)^{-1} = \sum_{k=0}^{m} N^k$$

**Proof**

Let $S = \sum_{k=0}^{m} N^k$.

Then

$$(I - N)S = \sum_{k=0}^{m} N^k - \sum_{k=1}^{m+1} N^k.$$

All intermediate terms cancel telescopically, leaving

$(I - N)S = I - N^{m+1}$.

Since

$N^{m+1} = 0,$

we obtain

$(I - N)S = I.$

Similarly,

$S(I - N) = I.$

Therefore,

$S = (I - N)^{-1}.$

**Theorem 11 (Exact Born Collapse Under Nilpotency)**

Let

$T = G_0(E)V$

be the Born transfer operator.

If

$T^{m+1} = 0, T^m \neq 0,$

then the exact solution of the Lippmann–Schwinger equation is

$$| \psi\rangle = \sum_{k=0}^{m} T^k \mid \phi\rangle.$$

This expression is exact, not asymptotic.

The truncation error is strictly zero.

**Proof**

By Proposition 4,

$I - T$

is invertible.

Applying Lemma 10 to

$N = T,$

we obtain

$$(I - T)^{-1} = \sum_{k=0}^{m} T^k$$

Since

$(I - T) \mid \psi\rangle = \mid \phi\rangle,$

it follows that

$$| \psi\rangle = (I - T)^{-1} \mid \phi\rangle = \sum_{k=0}^{m} T^k \mid \phi\rangle.$$

The truncation error is

$$R_m = \sum_{k=m+1}^{\infty} T^k \mid \phi\rangle.$$

But

$T^{m+1} = 0,$

hence every term in the remainder vanishes identically:

$R_m = 0.$

**Corollary 12 (Absence of a Smallness Condition)**

If

$T^{m+1} = 0,$

then the identity of Theorem 11 holds without any condition on

$\| T \|.$

**Proof**

The equality

$$(I - T)^{-1} = \sum_{k=0}^{m} T^k$$

is a purely algebraic identity whose validity depends only on the nilpotency condition

$T^{m+1} = 0,$

and not on the norm of $T$.

A nilpotent operator may have arbitrarily large norm. For example, if

$N^2 = 0,$

then

$(\lambda N)^2 = 0$

for every scalar

$\lambda,$

while

$\| \lambda N \| = | \lambda | \; \| N \|$

can be arbitrarily large.

**Corollary 13 (Decomposition of the Scattered State)**

Under the hypotheses of Theorem 11, the scattered state admits the exact decomposition

$| \psi \rangle = | \phi \rangle + T | \phi \rangle + T^2 | \phi \rangle + \cdots + T^m | \phi \rangle.$

Each term

$T^k | \phi \rangle$

collects the contribution of exactly $k$ successive interactions with the potential $V$.

**5. Interpretation Within the SON Framework**

The Unified Nilpotent Operational Framework (SON) systematically studies the exact termination regime of formal series

$$F(N) = \sum_{k=0}^{\infty} a_k N^k$$

when $N$ is nilpotent in an algebra $\mathcal{A}$.

The SON framework identifies three principal classes of instances:

1. **Formal series in algebras containing nilpotent elements:** evaluation of formal series on nilpotent elements of abstract algebras.
2. **Functions of nilpotent operators:** rational or analytic functions evaluated on nilpotent operators acting on operator spaces.
3. **Incidence algebras of acyclic quivers:** path algebras associated with directed acyclic graphs (DAGs).

The Born–SON system developed in this article belongs primarily to regime (ii), corresponding to the function

$$F(T) = (I - T)^{-1},$$

evaluated on the nilpotent transfer operator

$$T = G_0(E)V.$$

The connection with regime (iii) is equally direct: the transition graph $\Gamma_T$

introduced in Section 6 is precisely the quiver underlying the path algebra of the system.

**Definition 14 (Born–SON System)**

A Born–SON system is a quadruple

$$(\mathcal{H}, \mathcal{A}, T, m),$$

where:

- $\mathcal{H}$ is a finite-dimensional Hilbert space;
- $\mathcal{A}$ is an operator algebra;
- $T$ is the Born transfer operator;
- $m$ is the nilpotency degree:

$$T^{m+1} = 0, T^m \neq 0.$$

**Corollary 15 (Born–SON Collapse Principle)**

In every Born–SON system, the Born expansion collapses into the exact finite operational identity

$$(I - T)^{-1} = \sum_{k=0}^{m} T^k$$

The number of terms is exactly

$m + 1,$

equal to the nilpotency index of $T$.

**Comparison Between the Perturbative Born Regime and the Born–SON Regime**

| Feature | Perturbative Born Regime | Born–SON Regime |
|---|---|---|
| Series structure | Infinite | Finite ($m + 1$ terms) |
| Validity condition | $\parallel T \parallel < 1$ | $T^{m+1} = 0$ |
| Truncation error | Nonzero | Exactly zero |
| Nature of the mechanism | Analytic (convergence) | Algebraic (structure) |
| Physical origin | Weak potential | Acyclic transition graph |
| Smallness condition | Necessary | Unnecessary |

**6. Finite Acyclic Quantum Systems**

**Standing Assumption for Sections 6–13.**

From this point forward, we assume that

$\dim\ \mathcal{H} = n < \infty.$

All graph-theoretic and nilpotency arguments in the remainder of the paper are made under this finite-dimensional assumption.

**Definition 16 (Transition Graph)**

Let $\mathcal{B} = \{|\ 1\rangle, \dots, |\ n\rangle\}$ be an orthonormal basis of $\mathcal{H}$.

The transition graph associated with the transfer operator $T$ is the directed graph $\Gamma_T$,

defined as follows:

- the vertex set is

$V(\Gamma_T) = \{1, \dots, n\}$;

- there exists a directed edge

$i \to j$

if and only if

$\langle j \mid T \mid i \rangle \neq 0$.

Throughout the paper we use the standard column-vector convention. Thus,

$T_{ji} = \langle j|T|i$

denotes the amplitude for the transition

$|i\rangle \longrightarrow |j\rangle$.

Accordingly, a directed edge $i \to j$ in the transition graph corresponds to a nonzero matrix entry $T_{ji}$. For example, the entry $T_{21} = t_{21}$ represents the transition $1 \to 2$.

**Observation 17**

The index convention follows the physical direction of the transition:

the edge

$i \to j$

means that the system can transition from the state

$\mid i\rangle$

to the state

$\mid j\rangle$

under the action of the interaction operator.

**Definition 18 (Acyclic System)**

The system is said to be **acyclic** if the transition graph

$\Gamma_T$

is a directed acyclic graph (DAG), that is, if it contains no directed cycles.

**Theorem 19 (Acyclicity Implies Nilpotency)**

If $\Gamma_T$ is a DAG with $n$ vertices, then

$T^n = 0.$

**Proof**

Consider the matrix entry

$(T^k)_{ij}.$

By matrix multiplication,

$$(T^k)_{ij} = \sum_{r_1,\dots,r_{k-1}} T_{ir_{k-1}} T_{r_{k-1}r_{k-2}} \cdots T_{r_1 j}$$

A summand is nonzero only if the graph contains the directed chain

$j \to r_1 \to r_2 \to \cdots \to r_{k-1} \to i,$

that is, a directed path of length $k$ from $j$ to $i$.

Now suppose

$k = n.$

Any directed path of length $n$ in a graph with only $n$ vertices must repeat at least one vertex (pigeonhole principle).

Repeating a vertex in a directed path necessarily produces a directed cycle, contradicting the acyclicity assumption.

Therefore, no directed path of length $n$ exists in $\Gamma_T$.

Hence

$(T^n)_{ij} = 0$ for all $i, j$, which gives $T^n = 0$.

**Corollary 20 (Exact Born Expansion in Acyclic Systems)**

Let $\Gamma_T$ be a DAG with $n$ vertices and maximal directed-path length $m$, where

$m < n.$

Then

$T^{m+1} = 0,$

and the exact Born expansion contains exactly $m + 1$ terms:

$$|\psi\rangle = \sum_{k=0}^{m} T^k |\phi\rangle.$$

**Proof**

Because the graph contains a directed path of length $m$,

$T^m \neq 0.$

Since no path of length $m + 1$ exists,

$T^{m+1} = 0$

by the argument of Theorem 19.

Applying Theorem 11 yields the result.

**Proposition 21 (Born–SON Depth and Maximal Path Length)**

If $\Gamma_T$ is a DAG, then the nilpotency degree of $T$ coincides with the maximal directed-path length of the graph:

$m = \max\{\text{length of directed paths in } \Gamma_T\}.$

**Proof**

If the graph contains a directed path of length $m$, then at least one entry of

$T^m$

is nonzero, hence

$T^m \neq 0.$

If no directed path of length $m + 1$ exists, then every entry of

$T^{m+1}$

vanishes, implying

$T^{m+1} = 0.$

Therefore,

$m$

is precisely the nilpotency degree.

**Examples of Physical Acyclic Systems**

Several physically relevant quantum systems naturally produce acyclic transition graphs:

1. **Cascade spontaneous decay systems:** excited atoms decaying sequentially without reabsorption or reexcitation [1,12].

2. **Unidirectional tight-binding models (quantum ratchets):** charge transport under asymmetric dissipation.
3. **Ladder-type multilevel systems:** systems whose allowed transitions occur only toward lower-energy states.
4. **Open quantum Markov systems with irreversible dissipation:** systems in which dissipation breaks microscopic reversibility [11].
5. **Two-path quantum junctions:** mesoscopic transport devices in which carriers propagate through non-feedback paths [2].

**7. Introductory Example: Three-Level Cascade Atom**

We now present the simplest explicit example illustrating the Born–SON collapse mechanism.

Its limitations, discussed at the end of the section, motivate the richer and physically more significant diamond-graph example of Section 8.

**Description of the System**

Consider a three-level ladder-type atom with states

$|1\rangle, |2\rangle, |3\rangle$,

where:

- $|1\rangle$ is the ground state,
- $|2\rangle$ is the first excited state,
- $|3\rangle$ is the second excited state.

Allowed transitions occur only in the spontaneous-emission direction:

$|3\rangle \to |2\rangle \to |1\rangle$,

with transition amplitudes

$t_{21}, t_{32}$.

There is no reverse transition and no direct coupling between

$|3\rangle$

and

$|1\rangle$.

**Transition Graph**

The transition graph is the DAG

$1 \leftarrow 2 \leftarrow 3,$

with:

- 3vertices,
- maximal path length

$m = 2.$

**Transfer Operator**

In the ordered basis

$(|1\rangle, |2\rangle, |3\rangle),$

the transfer operator has matrix representation

$$T = \begin{pmatrix} 0 & t_{21} & 0 \\ 0 & 0 & t_{32} \\ 0 & 0 & 0 \end{pmatrix}.$$

**Verification of Nilpotency**

Direct computation gives

$$T^2 = \begin{pmatrix} 0 & 0 & t_{21}t_{32} \\ 0 & 0 & 0 \\ 0 & 0 & 0 \end{pmatrix},$$

and

$T^3 = 0.$

Thus,

$T$

is nilpotent of index 3, in agreement with Proposition 21.

**Exact Born Collapse**

By Theorem 11, the exact scattered state corresponding to the incoming state

$|\phi\rangle = |3\rangle$

is

$|\psi\rangle = (I + T + T^2)|3\rangle.$

We compute term by term:

$T|3\rangle = t_{32}|2\rangle,$

and

$T^2 \mid 3\rangle = t_{21} t_{32} \mid 1\rangle.$

Hence,

$\mid \psi\rangle = \mid 3\rangle + t_{32} \mid 2\rangle + t_{21} t_{32} \mid 1\rangle.$

**Born–SON Terms for the Three-Level Cascade Atom**

| Term | Physical Origin | Contribution |
|---|---|---|
| I | $I \mid 3\rangle$ | Free propagation |
| T | $T \mid 3\rangle$ | One transition |
| $T^2$ | $T^2 \mid 3\rangle$ | Two successive transitions |
| $T^3$ | $T^3 \mid 3\rangle$ | Impossible (acyclic graph) |

**Limitation of This Example**

The three-level cascade system clearly illustrates the collapse mechanism, but its nilpotency is immediately visible from the strictly triangular form of the matrix $T$.

Any matrix analyst would recognize the nilpotency without invoking the Born–SON framework.

The diamond-graph system of Section 8 is fundamentally richer:

- the nilpotency is no longer visually obvious,
- the system contains two interfering scattering paths,
- and the exact Born–SON collapse produces quantitatively meaningful physical effects:
    - constructive interference,
    - exact destructive interference,
    - and dark-state formation,

all of which are completely missed by the first-order Born approximation.

**8. Diamond Graph: Exact Interference and Dark States**

This section contains the principal result of the paper.

**Physical Motivation**

Whenever a quantum system provides two alternative routes connecting an initial state to a final state, the corresponding amplitudes combine algebraically.

The resulting total amplitude may be:

- larger than each individual contribution (constructive interference),
- smaller (partial destructive interference),
- or exactly zero (dark-state formation).

This phenomenon underlies several major physical effects, including:

- electromagnetically induced transparency (EIT) [3,4],
- Fano antiresonances,
- and dark states in quantum optics [13].

We now show that the Born–SON framework captures this interference mechanism exactly in a finite acyclic system with diamond-graph structure, whereas the first-order Born approximation fails completely.

**Description of the System**

Consider a four-level quantum system with basis states

$|1\rangle, |2\rangle, |3\rangle, |4\rangle,$

and allowed transitions

$|1\rangle \to |2\rangle, |1\rangle \to |3\rangle, |2\rangle \to |4\rangle, |3\rangle \to |4\rangle.$

There are:

- no reverse transitions,
- no couplings between the two branches,
- and no feedback loops.

The geometry of the transition graph is therefore the diamond graph:

Physical realizations include:

- Λ-type EIT configurations,
- mesoscopic two-path quantum junctions,
- and four-level systems in quantum optics.

**Transition Graph and Acyclicity**

The transition graph $\Gamma_T$ contains:

- vertices

$\{1,2,3,4\}$,

- directed edges

$1 \to 2, 1 \to 3, 2 \to 4, 3 \to 4.$

The graph is acyclic.

Indeed, the only directed paths are

$1 \to 2, 1 \to 3, 2 \to 4, 3 \to 4,$

together with the two paths

$1 \to 2 \to 4, 1 \to 3 \to 4.$

None of them forms a cycle.

The maximal path length is therefore

$m = 2.$

**Transfer Operator and Nilpotency**

In the ordered basis

$(|1\rangle, |2\rangle, |3\rangle, |4\rangle),$

the transfer operator is

$$T = \begin{pmatrix} 0 & 0 & 0 & 0 \\ t_{21} & 0 & 0 & 0 \\ t_{31} & 0 & 0 & 0 \\ 0 & t_{42} & t_{43} & 0 \end{pmatrix}.$$

Direct computation gives

$$T^2 = \begin{pmatrix} 0 & 0 & 0 & 0 \\ 0 & 0 & 0 & 0 \\ 0 & 0 & 0 & 0 \\ t_{42}t_{21} + t_{43}t_{31} & 0 & 0 & 0 \end{pmatrix}.$$

The only nonzero entry corresponds precisely to the two directed paths

$1 \to 2 \to 4, 1 \to 3 \to 4.$

Next,

$T^3 = 0,$

because the graph contains no directed path of length 3.

Thus,

$T$

is nilpotent of index 3.

**Observation 22 (Structural origin of nilpotency)**

In the three-level cascade, the strict triangularity of $T$ is an immediate consequence of the linear ordering of the states $3 \to 2 \to 1$. In the diamond graph, the strict lower triangularity of $T$ in the given basis $(|1\rangle, |2\rangle, |3\rangle, |4\rangle)$ is also visible. However, for a general acyclic system, the nilpotency of $T$ need not be visible from any particular matrix representation: it depends on the existence of a topological ordering of the vertices, which is guaranteed by acyclicity (Theorem 19) but which may not coincide with any physically motivated labeling. The graph-theoretic argument of Theorem 19 provides the correct structural criterion in full generality.

**Main Result**

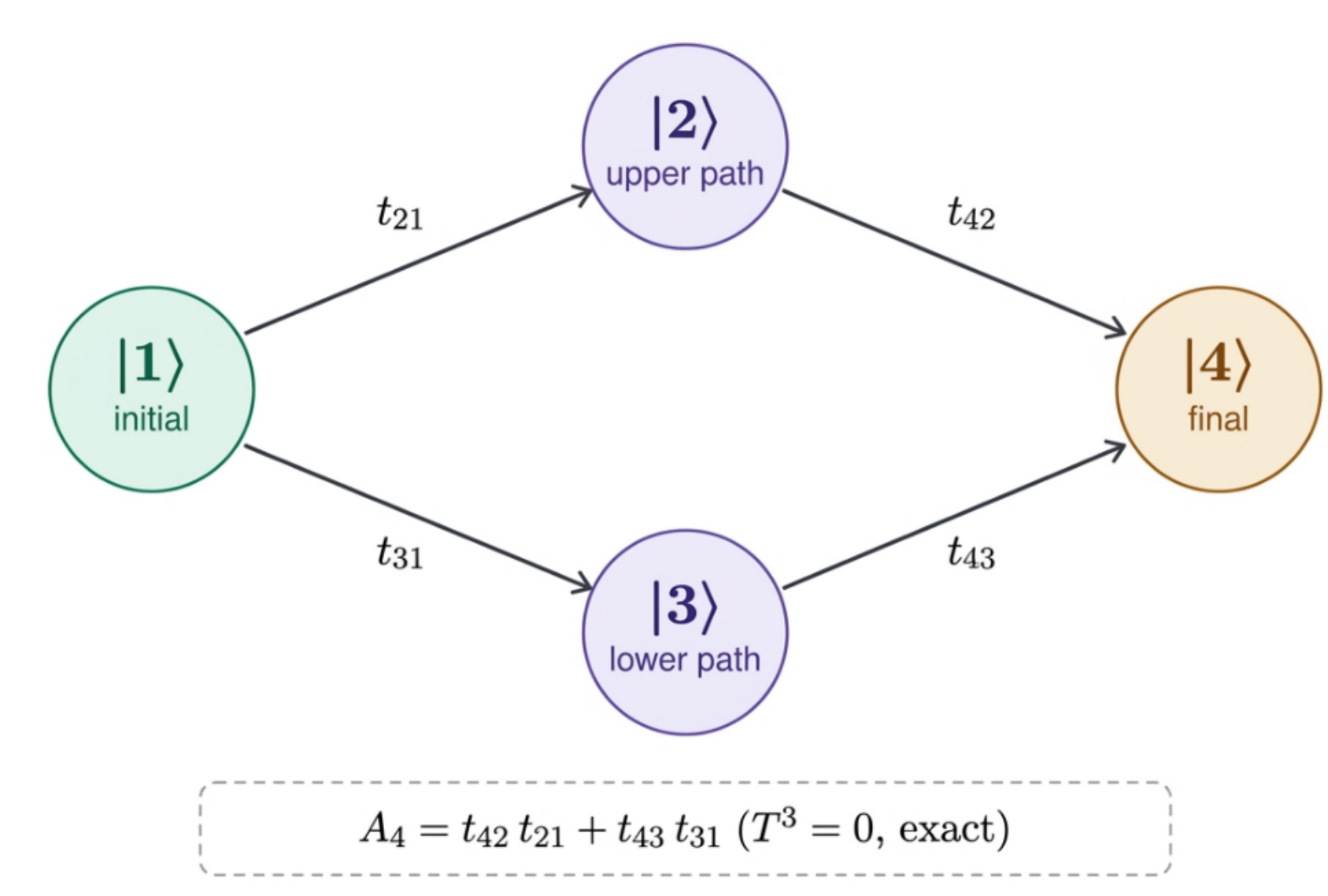

Figure 1. Diamond graph: four-level system with two interference paths. Arcs indicate allowed transitions with amplitudes $t_{21}, t_{31}, t_{42}, t_{43}$.

**Theorem 23 (Exact Interference in the Diamond Graph)**

Let $(\mathcal{H}, \mathcal{A}, T, m)$ be a Born–SON system with diamond-graph structure and transition amplitudes

$t_{21}, t_{31}, t_{42}, t_{43}$.

Then

$T^3 = 0,$

and the exact solution of the Lippmann–Schwinger equation with incoming state

$\mid \phi\rangle = \mid 1\rangle$

is

$\mid \psi\rangle = \mid 1\rangle + t_{21} \mid 2\rangle + t_{31} \mid 3\rangle + A_4 \mid 4\rangle,$

where the final-state interference amplitude is

$A_4 = t_{42}t_{21} + t_{43}t_{31}.$

This identity is exact.

The amplitude $A_4$ is precisely the algebraic sum of the two scattering paths of the diamond graph.

**Proof**

Since

$\Gamma_T$

is a DAG with maximal path length

$m = 2,$

Corollary 20 gives

$T^3 = 0.$

By Theorem 11,

$\mid \psi\rangle = (I + T + T^2) \mid 1\rangle.$

We compute:

$T \mid 1\rangle = t_{21} \mid 2\rangle + t_{31} \mid 3\rangle,$

and

$T^2 \mid 1\rangle = (t_{42}t_{21} + t_{43}t_{31}) \mid 4\rangle.$

Therefore,

$\mid \psi\rangle = \mid 1\rangle + t_{21} \mid 2\rangle + t_{31} \mid 3\rangle + (t_{42}t_{21} + t_{43}t_{31}) \mid 4\rangle.$

No higher-order term exists because

$T^3 = 0.$

**Observation 24 (Interpretation as a Feynman Path Sum)**

Equation

$A_4 = t_{42}t_{21} + t_{43}t_{31}$

is exactly the Feynman rule for two alternative quantum paths:

the total amplitude equals the sum of the amplitudes associated with each route.

The left branch

$1 \to 2 \to 4$

contributes

$t_{42}t_{21},$

while the right branch

$1 \to 3 \to 4$

contributes

$t_{43}t_{31}.$

Thus, the Born–SON collapse produces the exact algebraic form of quantum interference directly from nilpotency.

**Interference Regimes**

**Corollary 25 (Constructive Interference)**

If

$t_{42}t_{21} = t_{43}t_{31},$

then

$A_4 = 2t_{42}t_{21}.$

The amplitude at the final state is twice the contribution of a single path.

**Corollary 26 (Dark State: Exact Destructive Interference)**

If $t_{42}t_{21} = -t_{43}t_{31}$, then

$A_4 = 0.$

Hence

$| 4\rangle$

is a dark state:

the transition amplitude toward the final state vanishes exactly, independently of the magnitude of the individual couplings.

**Proof of Corollary 26**

Directly from

$A_4 = t_{42}t_{21} + t_{43}t_{31}$.

**Observation 27 (Nonperturbative Nature of the Dark State)**

The condition

$t_{42}t_{21} = -t_{43}t_{31}$

does not require any coupling amplitude to be small.

In particular, all couplings may be arbitrarily large.

The dark state is therefore a structural interference phenomenon rather than a weak-coupling effect.

**Interference Regimes in the Diamond Graph**

| **Regime** | **Condition** | **Final Amplitude $A_4$** | **Physical Effect** |
|---|---|---|---|
| Constructive interference | $t_{42}t_{21} = t_{43}t_{31}$ | $2t_{42}t_{21}$ | Amplification |
| Dark state | $t_{42}t_{21} = -t_{43}t_{31}$ | 0 exactly | Complete cancellation |
| Generic case | Arbitrary amplitudes | $t_{42}t_{21} + t_{43}t_{31}$ | Partial interference |

**Failure of the First-Order Born Approximation**

**Corollary 28 (Quantitative Failure of First-Order Born)**

The first-order Born approximation gives

$|\psi\rangle_{\text{Born-1}} = | 1\rangle + T\, | 1\rangle$.

Therefore,

$|\psi\rangle_{\text{Born-1}} = | 1\rangle + t_{21}\, | 2\rangle + t_{31}\, | 3\rangle$.

The predicted amplitude at the final state

$\mid 4\rangle$

is therefore

$A_4^{(\text{Born-1})} = 0,$

independently of the values of the transition amplitudes.

Hence the first-order Born approximation cannot distinguish:

- constructive interference,
- dark-state formation,
- or the generic interference regime.

It predicts the same null amplitude in all cases.

**Proof**

Since

$T \mid 1\rangle = t_{21} \mid 2\rangle + t_{31} \mid 3\rangle,$

there is no contribution along

$\mid 4\rangle$

at first order.

The only mechanism capable of reaching the final state is the second-order process

$T^2 \mid 1\rangle.$

**Corollary 29 (Non-perturbative nature of the first-order Born failure).**

For the diamond-graph system, the first-order Born approximation omits the only term that reaches the final state $|4\rangle$. Hence

$A_4^{(\text{Born-1})} = 0,$

whereas the exact amplitude is

$A_4 = t_{42}t_{21} + t_{43}t_{31}.$

Consequently, whenever $A_4 \neq 0$, the relative error is

$$\frac{\left|A_4 - A_4^{(\text{Born-1})}\right|}{|A_4|} = 1$$

This failure is structural rather than merely quantitative: it occurs because the first-order approximation cannot see paths of length two.

Proof.

By Corollary 28,

$A_4^{(\text{Born-1})} = 0.$

By Theorem 23,

$A_4 = t_{42}t_{21} + t_{43}t_{31}.$

If $A_4 \neq 0$, then

$\frac{\left|A_4 - A_4^{(\text{Born-1})}\right|}{|A_4|} = \frac{|A_4|}{|A_4|} = 1$

The missing contribution is precisely the second-order term $T^2|1\rangle$.

**Born–SON Terms for the Diamond Graph**

| Term | Origin | Final-State Contribution | Present in First-Order Born? |
|---|---|---|---|
| I | $I \mid 1\rangle$ | No interaction | |
| T | $T \mid 1\rangle$ | One transition | |
| $T^2$ | $T^2 \mid 1\rangle$ | Two transitions; interference | $A_4$ |
| $T^3$ | $T^3 \mid 1\rangle$ | Impossible, since $T^3 = 0$ | 0 |

**Connection with Established Physical Phenomena**

**Electromagnetically Induced Transparency**

In the Λ-configuration of electromagnetically induced transparency (EIT) [3,4], two field–matter coupling routes interfere destructively and generate a transparency window.

The dark-state condition of Corollary 26,

$t_{42}t_{21} = -t_{43}t_{31},$

is the exact algebraic analogue of that phenomenon.

It expresses transparency as a cancellation between two transition paths, without requiring the full Wigner–Weisskopf wave-function formalism.

**Fano Antiresonance**

In mesoscopic transport systems, the diamond graph is the minimal model of a two-path quantum junction.

The dark-state condition corresponds to the Fano antiresonance condition: the transmission amplitude vanishes when the two path amplitudes cancel exactly.

Corollary 28 shows that first-order Born theory cannot predict such an antiresonance, because it omits the second-order term responsible for the final-state transition.

**Dark States in Quantum Optics**

In three- and four-level quantum systems, dark states are superpositions of basis states that do not couple to the radiation field [14].

The Born–SON formulation identifies them algebraically as states for which the arrival amplitude satisfies

$A_4 = 0.$

Thus, the dark-state mechanism is encoded directly in the finite nilpotent structure of the transition graph.

**9. Full Resolvent and Transition Matrix**

**Full Resolvent**

The formal resolvent identity for the full Hamiltonian

$H = H_0 + V$

is

$G(E) = (E - H)^{-1}.$

In scattering theory one has, formally,

$G(E) = (I - G_0(E)V)^{-1}G_0(E).$

Since

$T = G_0(E)V,$

this becomes

$G(E) = (I - T)^{-1}G_0(E).$

**Theorem 30 (Exact Born–SON Resolvent)**

If $T^{m+1} = 0,$

then the full resolvent is exactly

$$G(E) = \left(\sum_{k=0}^{m} T^k\right) G_0(E).$$

**Proof**

By Lemma 10,

$$(I - T)^{-1} = \sum_{k=0}^{m} T^k$$

Multiplying on the right by

$G_0(E)$

gives

$$G(E) = (I - T)^{-1} G_0(E) = \left(\sum_{k=0}^{m} T^k\right) G_0(E).$$

**Transition Matrix**

The transition matrix, not to be confused with the transfer operator $T$, is defined by

$$\mathcal{T} = V + V G_0 V + V G_0 V G_0 V + \cdots .$$

Equivalently,

$$\mathcal{T} = V(I - G_0 V)^{-1}.$$

Since

$T = G_0 V,$

we may write

$$\mathcal{T} = V(I - T)^{-1}.$$

**Theorem 31 (Born–SON Collapse of the Transition Matrix ($\mathcal{T}$))**

If $T^{m+1} = 0,$

then the exact transition matrix is

$$\mathcal{T} = V \sum_{k=0}^{m} T^k.$$

**Proof**

By Lemma 10,

$$(I-T)^{-1}=\sum_{k=0}^{m}T^{k}$$

Multiplying on the left by

$V$

yields

$$\mathcal{T}=V(I-T)^{-1}=V\sum_{k=0}^{m}T^{k}.$$

**10. Error Control in the Quasi-Nilpotent Case**

In realistic physical systems, the transition graph may fail to be strictly acyclic.

This can occur, for example, because of:

- weak feedback loops,
- environmental back-action,
- weak reabsorption,
- or residual reverse transitions.

In such situations, the transfer operator $T$ is not exactly nilpotent, but it may be close to nilpotent in an effective sense.

**Definition 32 (Nilpotency Defect)**

The nilpotency defect of order $m+1$ is the operator

$$D_{m+1}(T)=T^{m+1}.$$

We say that $T$ is quasi-nilpotent of effective order $m+1$ if

$$\| T^{m+1} \| \ll 1.$$

**Proposition 33 (Exact Truncation Error in the Quasi-Nilpotent Case)**

Assume that the Neumann series of $T$ converges, for example under

$$\| T \| < 1.$$

If the series is truncated at order $m$, then the exact error satisfies

$$R_m=(I-T)^{-1}T^{m+1}\mid \phi\rangle.$$

**Proof**

The telescoping identity gives

$$(I-T)\sum_{k=0}^{m} T^k = I - T^{m+1}.$$

Multiplying on the left by

$(I-T)^{-1}$,

we obtain

$$\sum_{k=0}^{m} T^k = (I-T)^{-1}(I-T^{m+1}).$$

Therefore,

$$(I-T)^{-1} - \sum_{k=0}^{m} T^k = (I-T)^{-1}T^{m+1}$$

Applying both sides to

$\mid \phi\rangle$

gives

$R_m = (I-T)^{-1}T^{m+1} \mid \phi\rangle.$

**Corollary 34 (Norm Control of the Error)**

If $\parallel T \parallel < 1$,

then

$$\parallel R_m \parallel \leq \frac{\parallel T^{m+1} \parallel}{1 - \parallel T \parallel} \parallel \phi \parallel.$$

In the exact nilpotent case,

$T^{m+1} = 0$,

and therefore

$R_m = 0.$

**Proof**

From Proposition 33,

$$\| R_m \| \leq \| (I-T)^{-1} \| \quad \| T^{m+1} \| \quad \| \phi \|.$$

If

$$\| T \| < 1,$$

then

$$\| (I-T)^{-1} \| \leq \frac{1}{1-\| T \|}.$$

Thus,

$$\| R_m \| \leq \frac{\| T^{m+1} \|}{1-\| T \|} \quad \| \phi \|.$$

Example 35 (Numerical scale of the quasi-nilpotent error).

Suppose that the Neumann series converges and that

$$|T\| = 0.5, \quad |T^{m+1}\| = 0.01, \quad |\phi\| = 1.$$

By Corollary 34,

$$\|R_m\| \leq \frac{0.01}{1-0.5} = 0.02.$$

Thus, a small nilpotency defect produces a controlled truncation error.

In the exact Born–SON case,

$$T^{m+1} = 0,$$

and therefore

$$\|R_m\| = 0.$$

This illustrates the transition between the exact nilpotent regime and the perturbative quasi-nilpotent regime.

## 11. Born–SON Depth

**Definition 36 (Born–SON Depth)**

If $T^{m+1} = 0, T^m \neq 0,$

the Born–SON depth of the system is defined by

$$d_{\mathrm{BSON}}(T) = m.$$

Equivalently, in an acyclic transition graph,

$d_{\mathrm{BSON}}(T) = \max\{\text{length of directed paths in } \Gamma_T\}.$

Thus, the Born–SON depth measures the number of interaction layers required for the exact Born expansion to close.

**Computational interpretation.**

The Born–SON depth $d_{\mathrm{BSON}}(T) = m$ determines the number of sequential operator powers needed to compute the exact finite Born expansion

$\sum_{k=0}^{m} T^k.$

Thus, the exact Born–SON computation requires only the powers

$T, T^2, \dots, T^m,$

rather than the direct inversion of $I - T$. In dense matrix form, a general inversion of $I - T$ has cubic complexity $O(n^3)$, whereas the Born–SON procedure exploits the structural fact that all powers beyond $m$ vanish. For sparse transition graphs, this can be substantially cheaper, since the computation follows only the directed paths of the acyclic graph.

**Example 37**

1. **Three-level cascade system:**

$d_{\mathrm{BSON}} = 2,$

corresponding to the path

$3 \to 2 \to 1.$

2. **Diamond graph:**

$d_{\mathrm{BSON}} = 2,$

corresponding to the two paths

$1 \to 2 \to 4, 1 \to 3 \to 4.$

3. **$n$-level cascade:**

$1 \leftarrow 2 \leftarrow 3 \leftarrow \cdots \leftarrow n.$

In this case,

$d_{\mathrm{BSON}} = n - 1.$

4. **Double diamond graph:** If the system contains two successive branching-and-recombination stages, the maximal directed-path length is 4, hence

$d_{\mathrm{BSON}} = 4.$

## 12. Scope and Limitations

### What This Article Establishes

This paper proves the following points:

1. The Born series is formally the Neumann series of the transfer operator

$T = G_0(E)V.$

2. If

$T^{m+1} = 0,$

the Born series terminates exactly after

$m + 1$

terms, with no condition on

$\| T \|$

and with exactly zero truncation error.

3. The acyclicity of the transition graph

$\Gamma_T$ implies nilpotency of $T,$

with nilpotency degree equal to the maximal path length of the graph.

4. In the four-level diamond graph, the Born–SON series captures exactly the interference between two scattering paths.
5. The dark-state condition is obtained algebraically as

$t_{42}t_{21} = -t_{43}t_{31}.$

6. First-order Born theory predicts zero final-state amplitude in all regimes and therefore cannot distinguish constructive interference, destructive interference, or the generic case.
7. The SON framework provides a structural algebraic interpretation of this exact finite regime.

### What This Article Does Not Claim

1. We do not claim that every Born series is nilpotent.

Nilpotency of $T$ requires acyclicity of the transition graph, which is a structural condition on the potential and resolvent.

2. We do **not** claim that the SON framework replaces standard scattering theory.

The Born–SON regime is complementary to the perturbative Born regime.

3. We do not claim that the nilpotent Neumann identity is new as a result in linear algebra. The algebraic identity $(I - N)^{-1} = \Sigma N^{\mathrm{k}}$ for nilpotent $N$ is classical; see, for instance, [10,16,17] for treatments in the context of scattering and spectral theory.The contribution lies in identifying a natural physical class of systems where this identity becomes operationally relevant and produces quantitative consequences not accessible to first-order perturbation theory.

4. We do not claim that the classical Born approximation is incorrect in the regime for which it was designed.

The classical Born approximation remains appropriate for weak-potential perturbative systems.

**13. Conclusions**

The Born series can be understood as the Neumann series of the transfer operator

$T = G_0(E)V.$

In the classical perturbative regime,

$\| T \| < 1,$

this expansion is infinite, and each term of order $k$ is interpreted as a progressively smaller correction.

In the Born–SON regime, by contrast, when

$T^{m+1} = 0,$

the expansion collapses into the exact algebraic identity

$$(I - T)^{-1} = \sum_{k=0}^{m} T^k$$

The sufficient structural condition for this regime is the acyclicity of the transition graph

$\Gamma_T.$

If the allowed transitions of the system contain no directed cycles, then $T$ is automatically nilpotent and the Born series terminates exactly.

The most significant result is Theorem 23:

in the four-level system with diamond-graph structure, the final-state transition amplitude is

$A_4 = t_{42}t_{21} + t_{43}t_{31},$

the exact sum of the two path amplitudes in the diamond graph.

This algebraic identity encodes the three fundamental interference regimes:

- constructive interference,
- exact destructive interference,
- and the generic partial-interference regime.

First-order Born approximation predicts

$A_4^{(\mathrm{Born}-1)} = 0$

in every case, thereby failing quantitatively across all regimes.

The conceptual chain of the paper is:

$$\text{Acyclic transition graph} \Longrightarrow \text{nilpotent transfer operator} \Longrightarrow \text{finite Neumann identity} \Longrightarrow \text{exact Born collapse} \Longrightarrow \text{exact quantum interference.}$$

The SON framework provides the unified algebraic language that organizes this regime.

The present article is therefore an instance of SON in quantum scattering theory, with concrete and verifiable physical consequences.

**Acknowledgments.** The author thanks the School of Mathematics of the Universidad Autónoma de Santo Domingo (UASD) for institutional support. This work received no specific external funding.

**Appendix A. Exact Algebraic Collapse (Self-Contained)**

This appendix proves, in a fully self-contained manner, the specific algebraic collapse used throughout the paper. It is a concrete instance of the general SON Collapse Lemma developed in [7], applied to the function

$F(T) = (I - T)^{-1}$.

However, no result from [7] is needed to verify the arguments below.

This appendix presents the three algebraic results upon which the paper is based.

**Lemma A.1 (Determinant of $I - N$ for Nilpotent $N$)**

Let $N^{m+1} = 0$

on a finite-dimensional vector space.

Then

$\det(I - N) = 1.$

In particular,

$I - N$

is invertible.

**Proof**

Since

$N^{m+1} = 0,$

the operator $N$ is nilpotent.

Hence all eigenvalues of $N$ are equal to zero.

Therefore, all eigenvalues of

$I - N$

are equal to one.

Because the determinant is the product of the eigenvalues,

$\det(I - N) = 1.$

**Lemma A.2 (Exact Telescoping Identity)**

Let $N$ be an operator satisfying

$N^{m+1} = 0.$

Then

$$(I - N)\sum_{k=0}^{m} N^k = I.$$

Equivalently,

$$(I - N)^{-1} = \sum_{k=0}^{m} N^k$$

**Proof**

Compute directly:

$$(I - N)\sum_{k=0}^{m} N^k = \sum_{k=0}^{m} N^k - \sum_{k=1}^{m+1} N^k$$

All intermediate terms cancel telescopically:

$N^1, \dots, N^m$.

The only surviving terms are

$I - N^{m+1}$.

Since

$N^{m+1} = 0$,

we obtain

$$(I - N)\sum_{k=0}^{m} N^k = I.$$

Thus,

$$(I - N)^{-1} = \sum_{k=0}^{m} N^k$$

**Corollary A.3 (Exact Born Collapse Under Nilpotency)**

Let $T = G_0(E)V$ be the Born transfer operator.

If $T^{m+1} = 0$,

then

$$(I - T)^{-1} = \sum_{k=0}^{m} T^k$$

is an exact algebraic identity.

Consequently, the solution of the Lippmann–Schwinger equation is

$$\mid \psi\rangle = \sum_{k=0}^{m} T^k \mid \phi\rangle,$$

with strictly zero truncation error and no condition on

$\parallel T \parallel$.

**Proof**

By Lemma A.1,

$I - T$

is invertible.

By Lemma A.2,

$$(I - T)^{-1} = \sum_{k=0}^{m} T^k$$

Applying this operator to

$$| \phi\rangle$$

gives

$$| \psi\rangle = \sum_{k=0}^{m} T^k | \phi\rangle.$$

The identity is algebraically exact because the series terminates through nilpotency,

$$T^{m+1} = 0,$$

rather than through convergence arguments.

**Remark A.4 (Formal verification).** The algebraic identities of Lemmas A.1–A.2 and Corollary A.3 have been formally verified in Lean 4 using Mathlib. The verification code is available at https://doi.org/10.5281/zenodo.20113599